%% file: proceedings_freiss.tex
\documentclass[10pt, paper=a4, UKenglish]{article}
\usepackage{graphicx}
\usepackage{enumitem}
\def\Title#1{\begin{center} {\Large #1 } \end{center}}
\def\Author#1{\begin{center}{ \sc #1} \end{center}}
\def\Address#1{\begin{center}{ \it #1} \end{center}}

\newcommand\pubblock{\rightline{\begin{tabular}{l} Proceedings of the CTD/WIT 2019\\ \pubnumber\\
         \pubdate  \end{tabular}}}

\newenvironment{Abstract}{\begin{quotation} \begin{center} 
             \large ABSTRACT \end{center}\bigskip 
      \begin{large}}{\end{large} \end{quotation}}

\newenvironment{Presented}{\begin{quotation} \begin{center} 
             PRESENTED AT\end{center}\bigskip 
      \begin{center}\begin{large}}{\end{large}\end{center} \end{quotation}}

\def\Acknowledgements{\bigskip  \bigskip \begin{center} \begin{large}
      \bf ACKNOWLEDGEMENTS \end{large}\end{center}}

\input econfmacros.tex

\usepackage{color}
\usepackage{lineno}
\usepackage{subcaption}
\usepackage{hyperref}
 \usepackage{ragged2e}

\newcommand\pubnumber{PROC-CTD19-129}

\newcommand\pubdate{\today}

\def\affiliation{
LPNHE, Sorbonne Universit\'{e}, Paris Diderot Sorbonne Paris Cit\'{e}, CNRS/IN2P3, Paris,  \\
France}

\newcommand{\conference}{Connecting the Dots and Workshop on Intelligent Trackers (CTD/WIT 2019)\\
Instituto de F\'isica Corpuscular (IFIC), Valencia, Spain\\ 
April 2-5, 2019}

\usepackage{fancyhdr}
\definecolor{mygrey}{RGB}{105,105,105}
\newcommand{\myhref}[2]{
	\href{#1}{\textcolor{blue}{#2}}
}

\begin{document}


\large
\begin{titlepage}
\pubblock

\vfill
\Title{Report from RAMP challenge on fast vertexing}
\vfill

\Author{Florian Reiss}
\Address{\affiliation}
\vfill

\begin{Abstract}
{\justify
Reconstructing the vertices of primary interactions at the LHCb experiment is an essential part of its online data acquisition sequence. The quest for ever higher rates and luminosities gives raise to new challenges for such algorithms. The LHCb Upgrade detector will start taking data in 2021 at an instantaneous luminosity of $\mathcal{L} = 2 \times 10^{33} cm^{-2} s^{-1}$. A RAMP challenge on fast primary vertex reconstruction using simulated data under such conditions was performed at the RAPID2018 workshop held in November 2018 in Dortmund with the aim of exploring new techniques and ideas to solve this problem.	
}
\end{Abstract}

\vfill

\begin{Presented}
\conference
\end{Presented}
\vfill
\end{titlepage}
\def\thefootnote{\fnsymbol{footnote}}
\setcounter{footnote}{0}
%

\normalsize 


\section{Introduction}
\label{intro}
The reconstruction of vertices plays a crucial role in many collider experiments. Going to collisions at ever higher rate and luminosity challenges traditional algorithms: They might be too slow or unable to reach good performance at environments with larger pile-up. Novel techniques might be required. Of current great interest is to approach such problems with machine learning algorithms. Since there is usually not a "best" solution deductible from first principles, developing such algorithms lends itself for working collaboratively or having the problem looked at from different angles. This is usually done with collaborative and/or competitive approaches.

The RAMP (Rapid analytics and model prototyping) framework \cite{ref:ramp} provides a platform for organizing data challenges with the goal of combining collaboration with competition. It does so by first having a 'closed' phase, where participants can compare their solutions with each other based on a scoreboard, but can not see the code itself. This is then followed up by an 'open' phase, where all submitted code is visible to everybody encouraging the sharing of ideas and trying out and improving on other approaches.

A challenge using the RAMP framework has been carried out in November 2018 \footnote{\myhref{https://rapid2018.org/}{https://rapid2018.org/}} with the context of finding primary vertices using simulated data of the LHCb Upgrade detector.

\section{Primary vertices at LHCb}
\label{section}
The LHCb experiment \cite{ref:lhcb-det} is a forward-arm spectrometer designed for studying beauty and charm mesons produced in proton-proton collisions at the Large Hadron Collider. Its data acquisition deciding which data to keep (trigger decision) relies on the signature of long-lived hadrons, which decay at a measurable distance from the interaction region. For Run3, beginning in 2021, the LHCb detector will be upgraded and move to a fully software-based trigger consisting of two stages \cite{ref:lhcb-upgrade}. The first stage (HLT1) needs to reconstruct events at a rate of 30 MHz.

After tracks -- the traces left by the particles -- are reconstructed in the vertex locator (Velo), the sub-detector closest to the interaction region, they can be used to find primary vertices (PVs), needed to identify displaced signatures. Displaced signatures are typically identified by tracks having a large impact parameter IP with respect to each PV. The IP is the transverse distance of closest approach of a track to a vertex. For an even better discrimination, the IP significance $\chi^2_{IP}$, which can be understood as the IP divided by its uncertainty, is commonly used.

\subsection{Performance parameters}
In terms of physics performance, the following parameters are important to achieve the physics goals of the LHCb experiment. They can be studied using Monte-Carlo simulation where the properties of the simulated true PVs are known.
\begin{description}[leftmargin=0cm]
\item[Efficiency] The PV finding efficiency is the number of reconstructed PVs matched to a true PV divided by the total number of reconstructible true PVs. A high efficiency is needed to ensure that the minimum IP of a track w.r.t each vertex can be properly determined. 
	\item[Fake rate] The fake rate is the number of reconstructed PVs not matched to a true PV divided by the total number of reconstructed PVs. A too large fake rate could distort the impact parameter of a track. In addition, a secondary vertex from a long-lived particle could be wrongly reconstructed as fake PV, reducing the efficiency of that signal.
	\item[Position resolution] The resolution is defined as the standard deviation of the distribution of the difference in (x,y,z)-position between a true PV and its matched reconstructed PV. The resolution influences what precision and accuracy can be reached by the impact parameter and $\chi^2_{IP}$.
\end{description}

\section{Setting up the challenge}
\begin{figure}[t]
	\begin{subfigure}[t]{0.45\textwidth}
		\includegraphics[width=\textwidth]{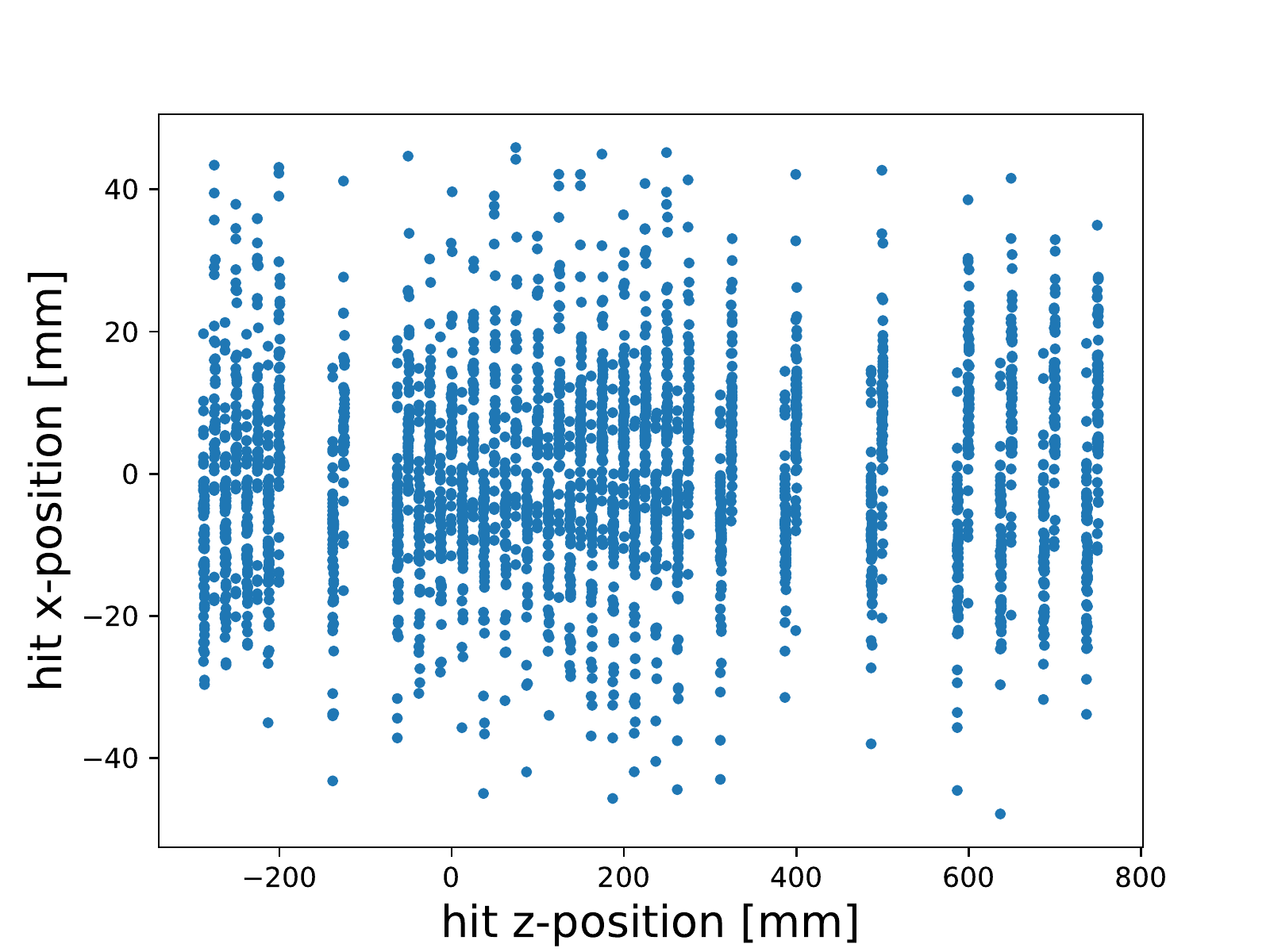}
	\end{subfigure}
	\begin{subfigure}[t]{0.45\textwidth}
		\includegraphics[width=\textwidth]{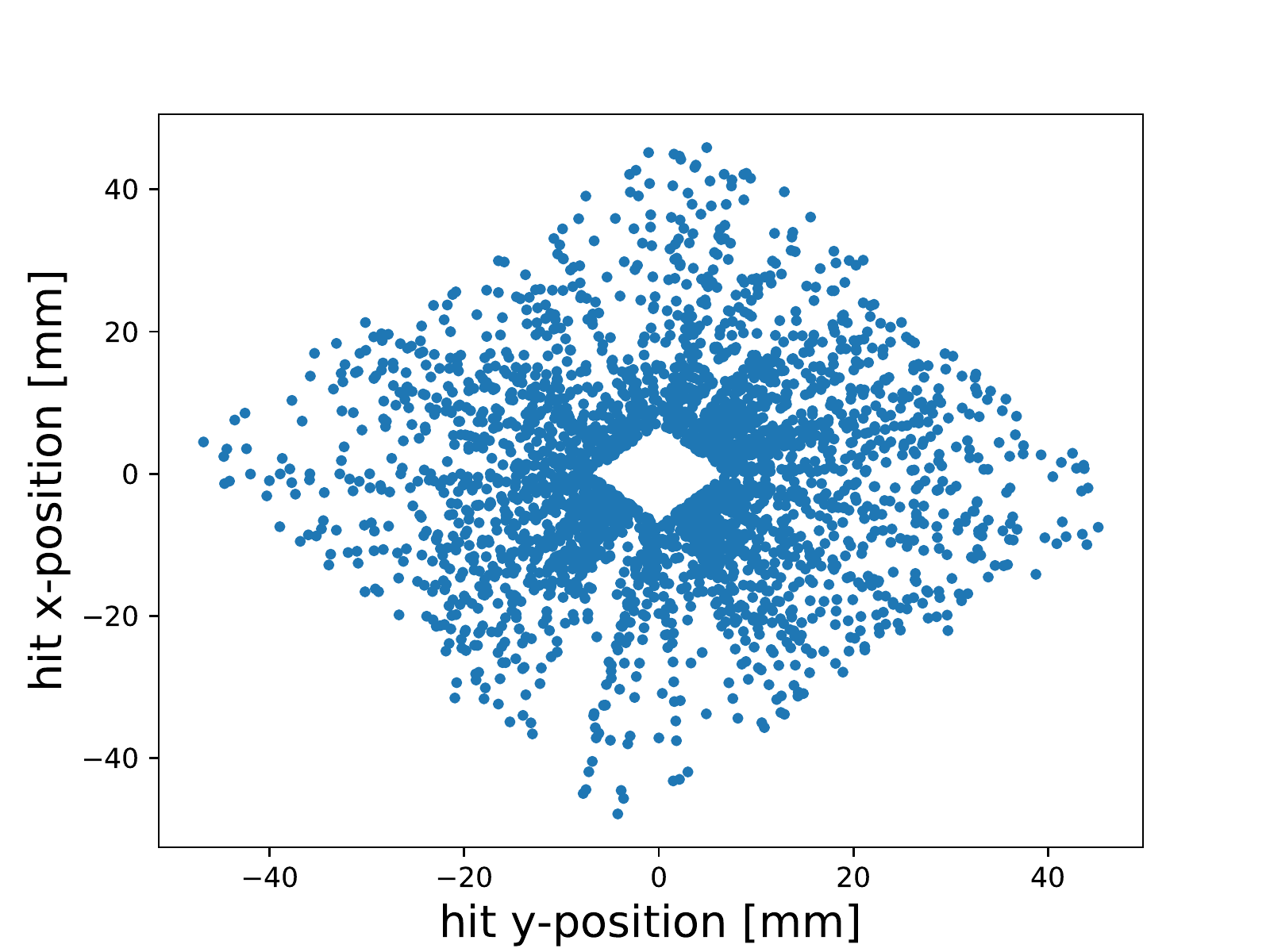}
	\end{subfigure}
\caption{Distribution of hits in the Vertex Locator for a typical event.}
\label{fig:velo_hits}
\end{figure}

The challenge uses simulated LHCb data under Run3 conditions, consisting of events with 6 PVs and about 200 tracks in the vertex locator on average.
The provided training and testing data consists of the (x,y,z)-positions of the hits in the Velo and the track state and covariance matrix of each reconstructed Velo track, extrapolated to the point of closest approach to the beamline. The track state and covariance matrix are calculated with a simplified Kalman filter, which treats the x- and y-components independently. Consequently, x and y are uncorrelated and the corresponding elements in the covariance matrix are zero. Additionally the error calculation is simplified, resulting in the uncertainty on the x-position (x-slope) to be exactly the same as the uncertainty on the y-position (y-slope).
The distribution of hits in the Velo for one event is shown in Figure~\ref{fig:velo_hits}. The coordinate system uses the beamline as z-axis.

As underlying truth the position and number of tracks of the simulated PVs are provided.
The solutions are required to produce a list of the positions of reconstructed PVs for each event. To keep matters more simple, no determination of uncertainties is required. To obtain the physics performance parameters, the list of reconstructed PVs is matched to the list of truth PVs for each event. A MC PV is considered as matched if the closest reconstructed PV is within a certain interval around its z-position. A reconstructed PV is only allowed to be matched to a single MC PV.

To get started with the challenge and to have a baseline for participants to compare their solutions to, a basic algorithm for PV reconstruction is provided. It should be noted that it is not optimized for Run3 conditions and does not include any fake removal, resulting in an efficiency of about 90\% and a fake rate of about 10\%; worse than what is necessary for the use case at LHCb, giving the participants an example to improve on.


\subsection{Defining the score}

\begin{figure}[t]
	\centering
	\includegraphics[height=7cm]{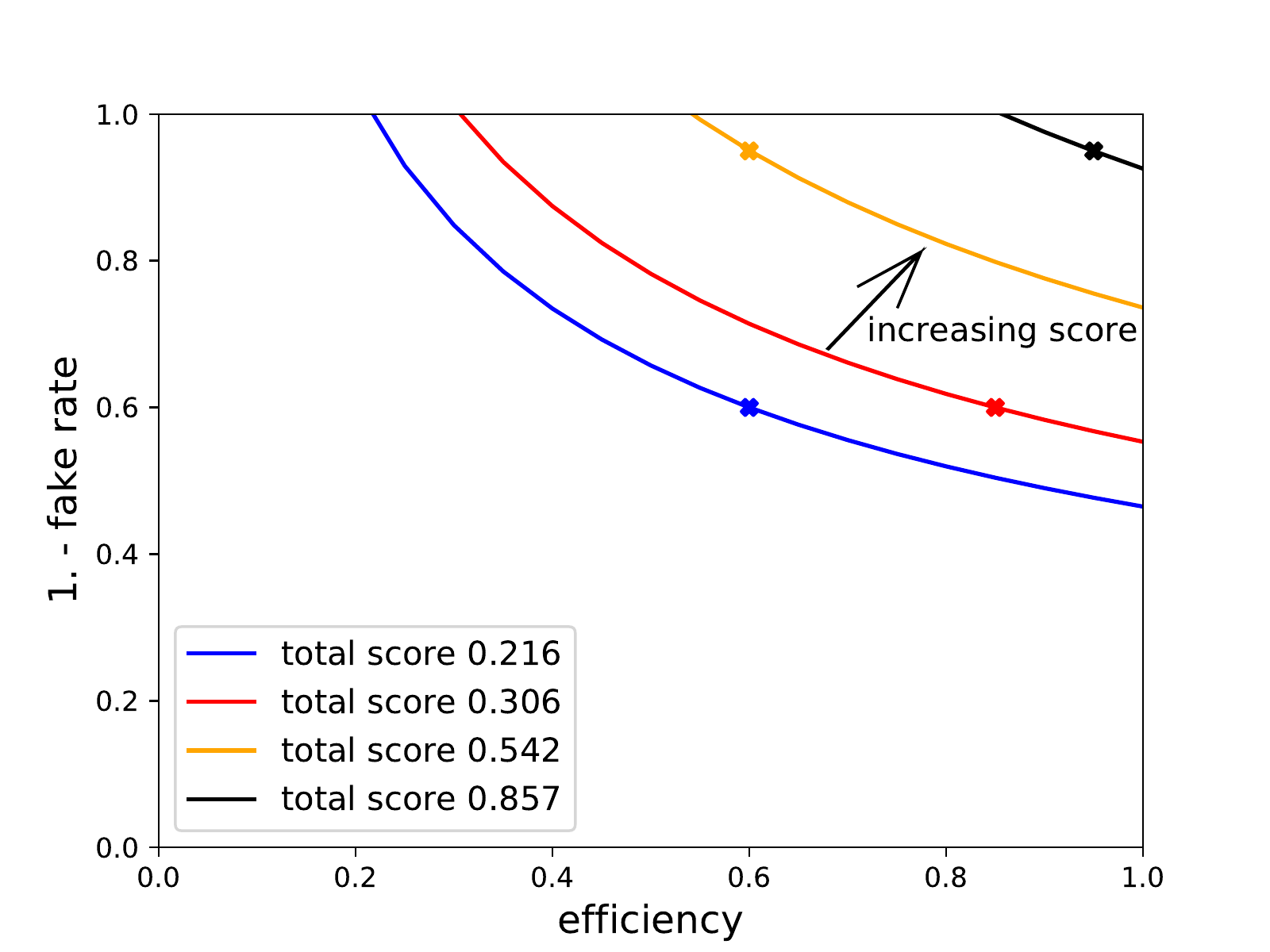}
	\caption{Illustration of total scores at different efficiencies and fake rates, assuming identical resolutions.}
	\label{fig:scoring}
\end{figure}

An algorithm for reconstructing PVs has to fulfill various requirements: It should have a high efficiency $\epsilon$, low fake rate $f$ and small resolutions $\sigma_{x,y,z}$ of the PV position, while being reasonably fast. Encapsulating all this in one single score for a challenge turns out to be no simple task. For the purpose of the workshop the various parameters are combined as
\begin{equation}
	 \epsilon \cdot (1- f)^2 \cdot \frac{N}{\sigma_x \sigma_y \sigma_z},
\end{equation} 
with a normalization constant $N$ chosen to obtain scores in the order of $\mathcal{O}(1)$. The execution time of the solutions is not considered. Using this definition means that solutions with vastly different physics performances can still obtain the same score as illustrated in Figure~\ref{fig:scoring}, where lines of equal scores are drawn at different $\epsilon$ and $f$, but fixed $\sigma_{x,y,z}$. Having low efficiency and low fake rate or high efficiency and high fake rate can result in the same score, with solutions improving on both increasing the score.

\section{Analyzing the submissions}
\begin{figure}[t]
	\centering
	\includegraphics[height=8cm]{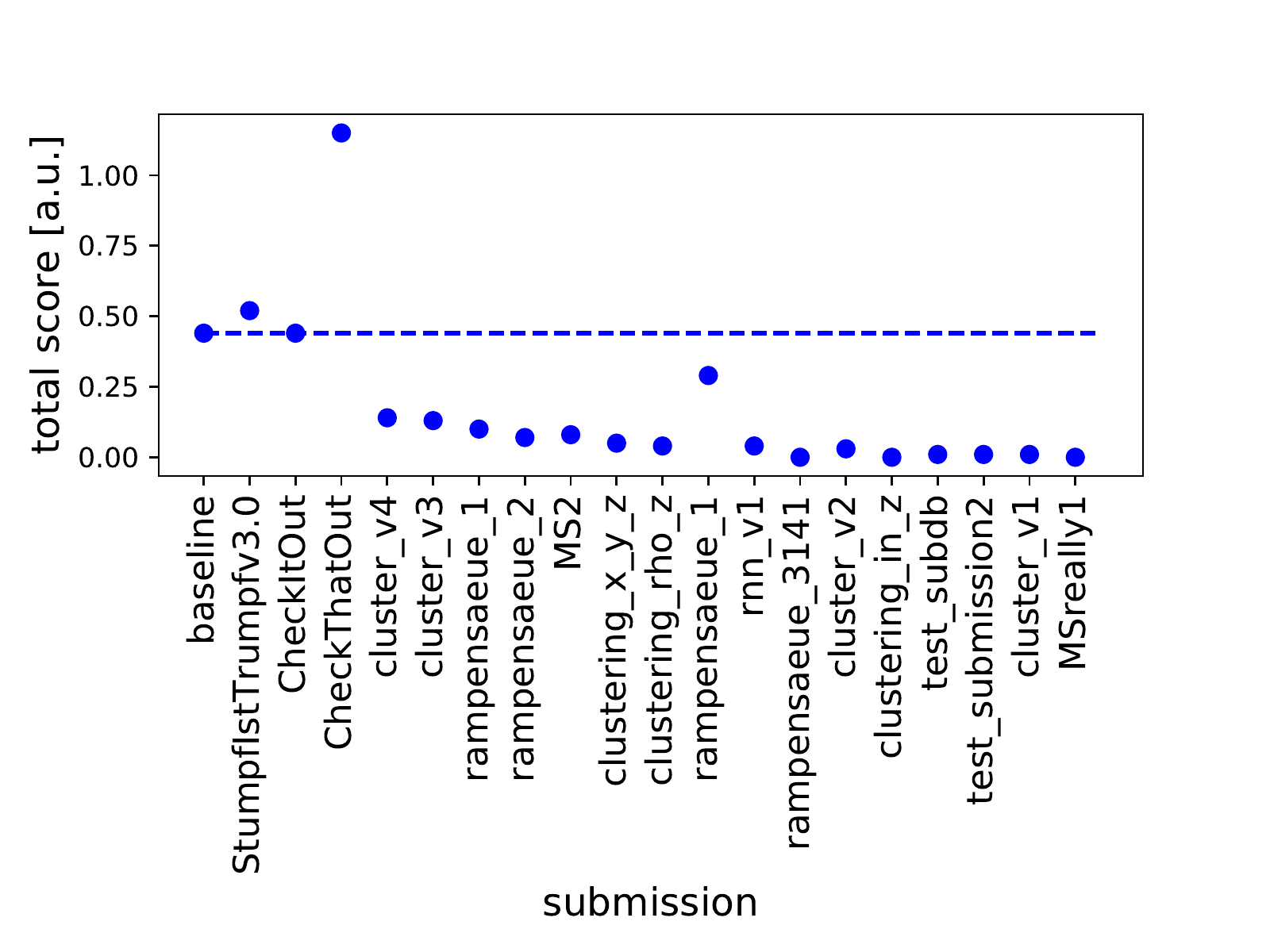}
	\caption{Comparison of the total scores of the submissions. The baseline is shown as dashed line.}
	\label{fig:tot_score}
\end{figure}

\begin{figure}[t]
	\centering
		\begin{subfigure}[t]{0.45\textwidth}
		\includegraphics[width=\textwidth]{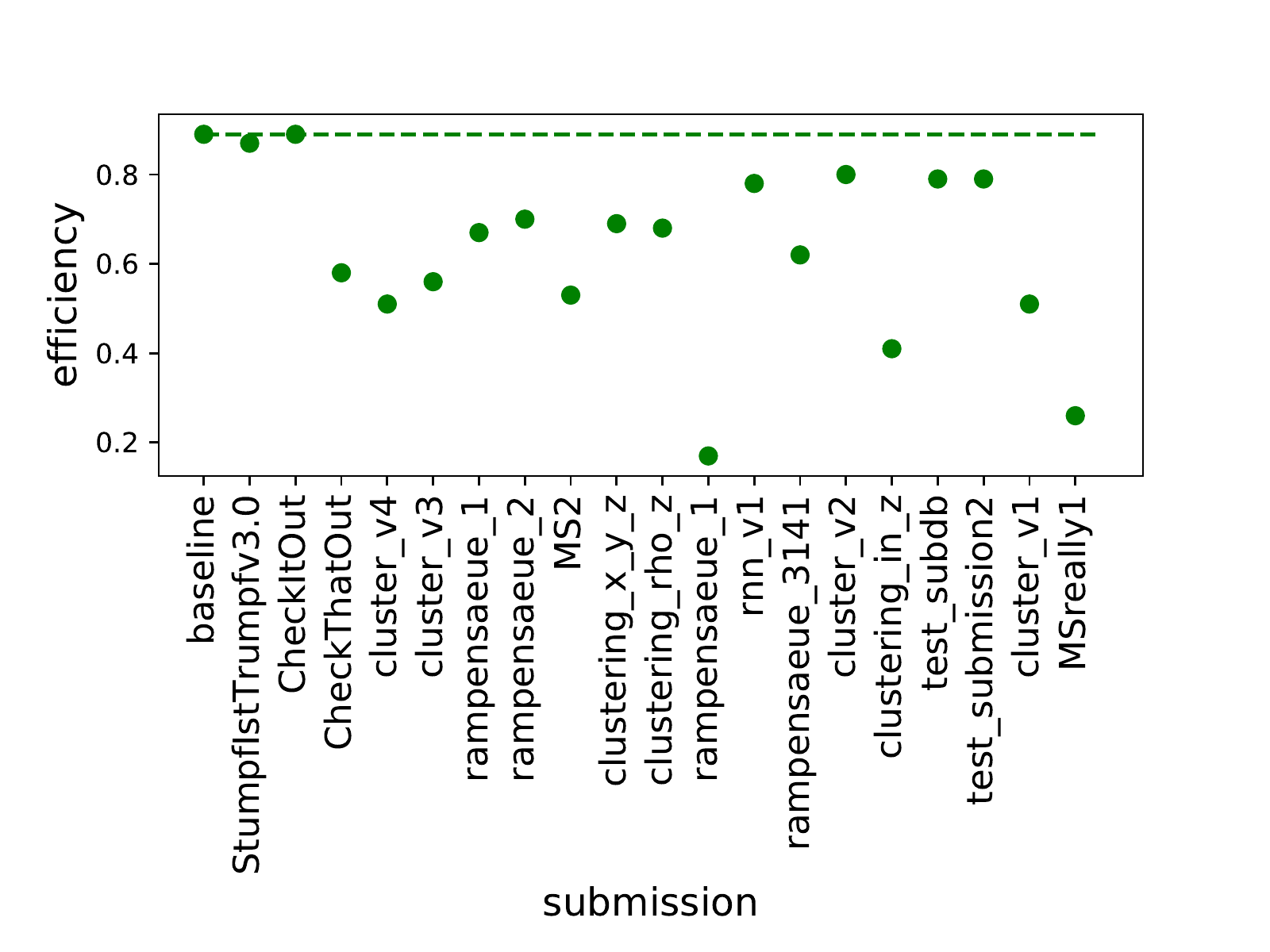}
	\end{subfigure}
		\begin{subfigure}[t]{0.45\textwidth}
	\includegraphics[width=\textwidth]{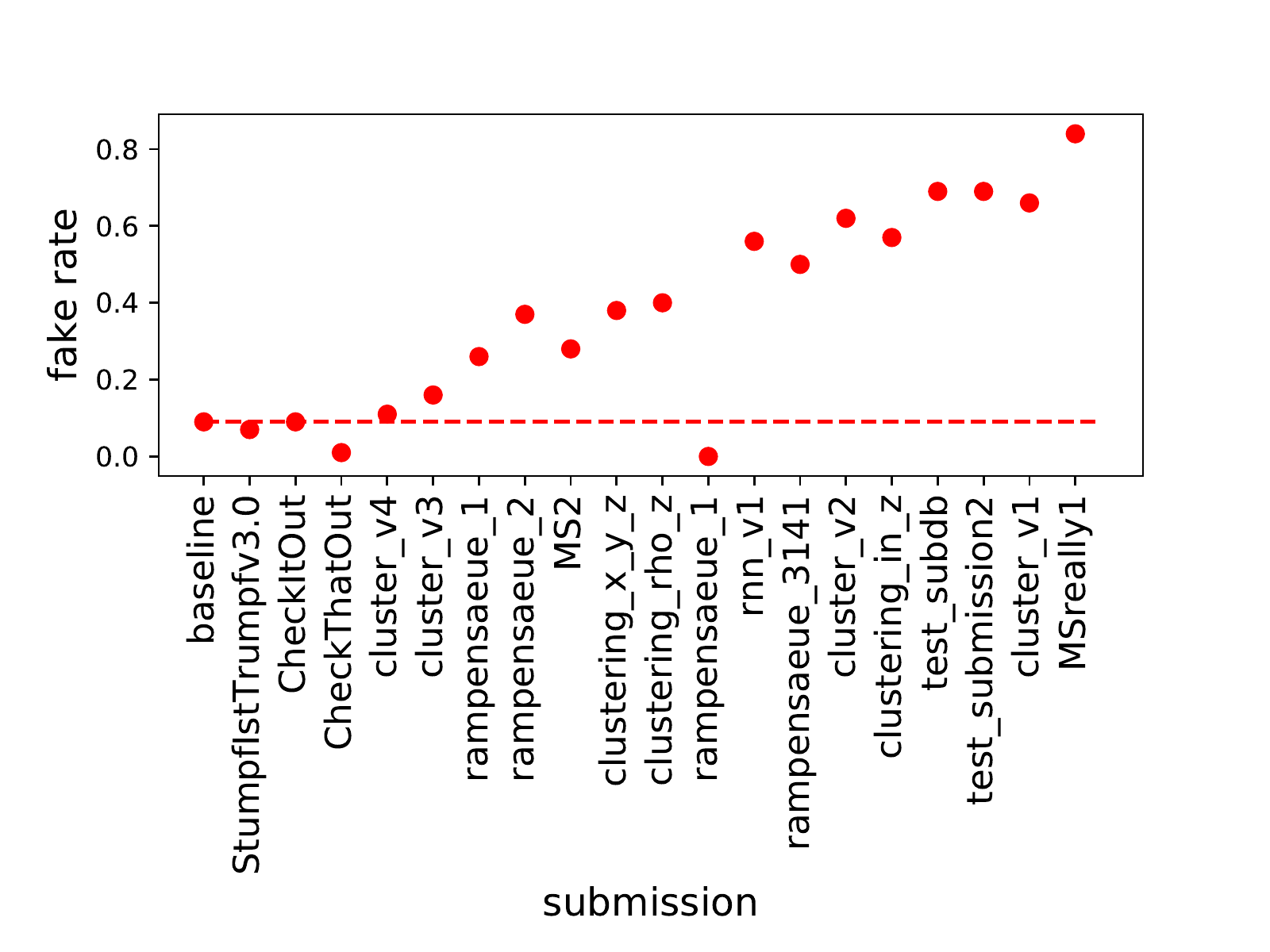}
\end{subfigure}

		\begin{subfigure}[t]{0.45\textwidth}
	\includegraphics[width=\textwidth]{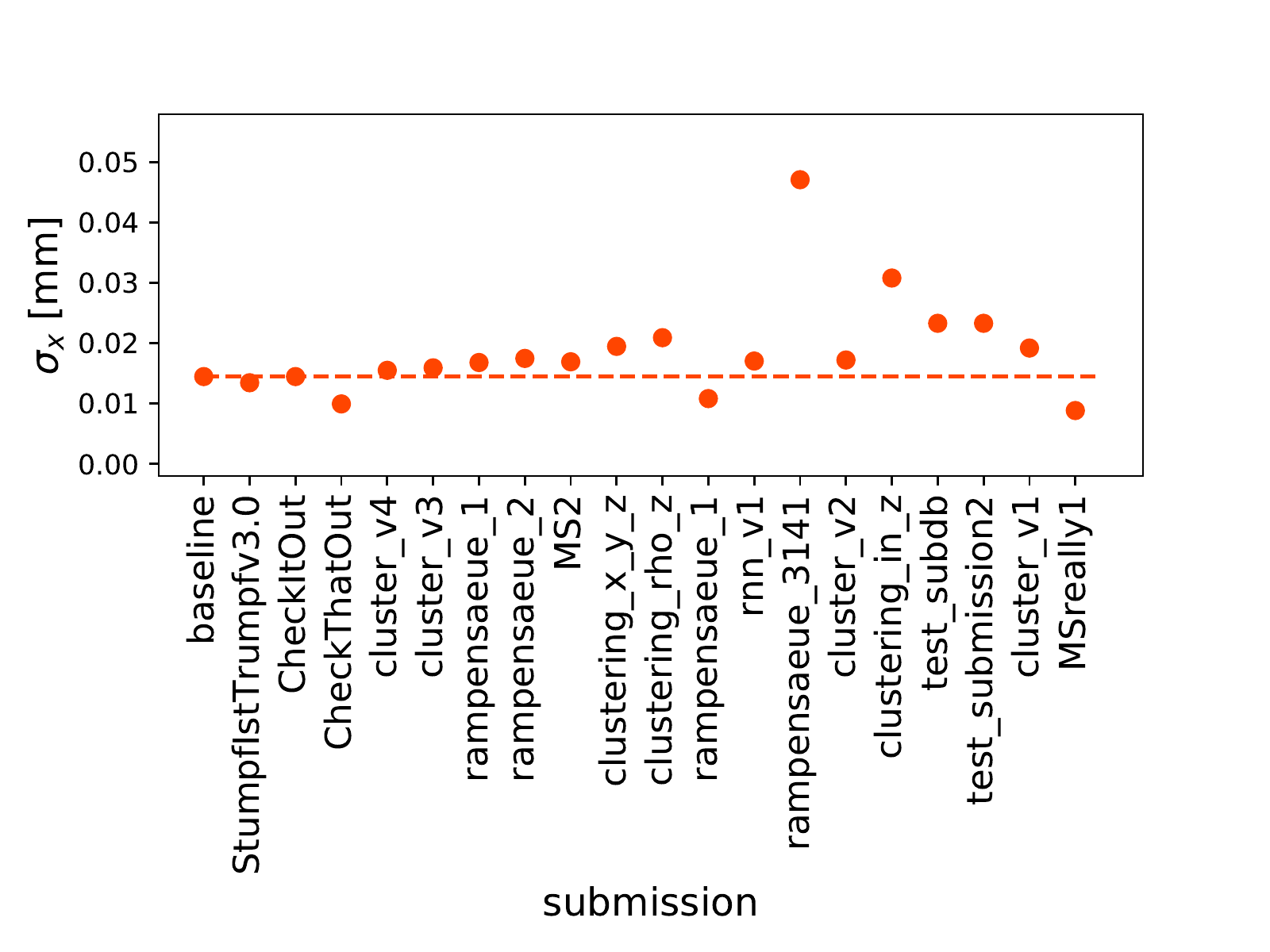}
\end{subfigure}
		\begin{subfigure}[t]{0.45\textwidth}
	\includegraphics[width=\textwidth]{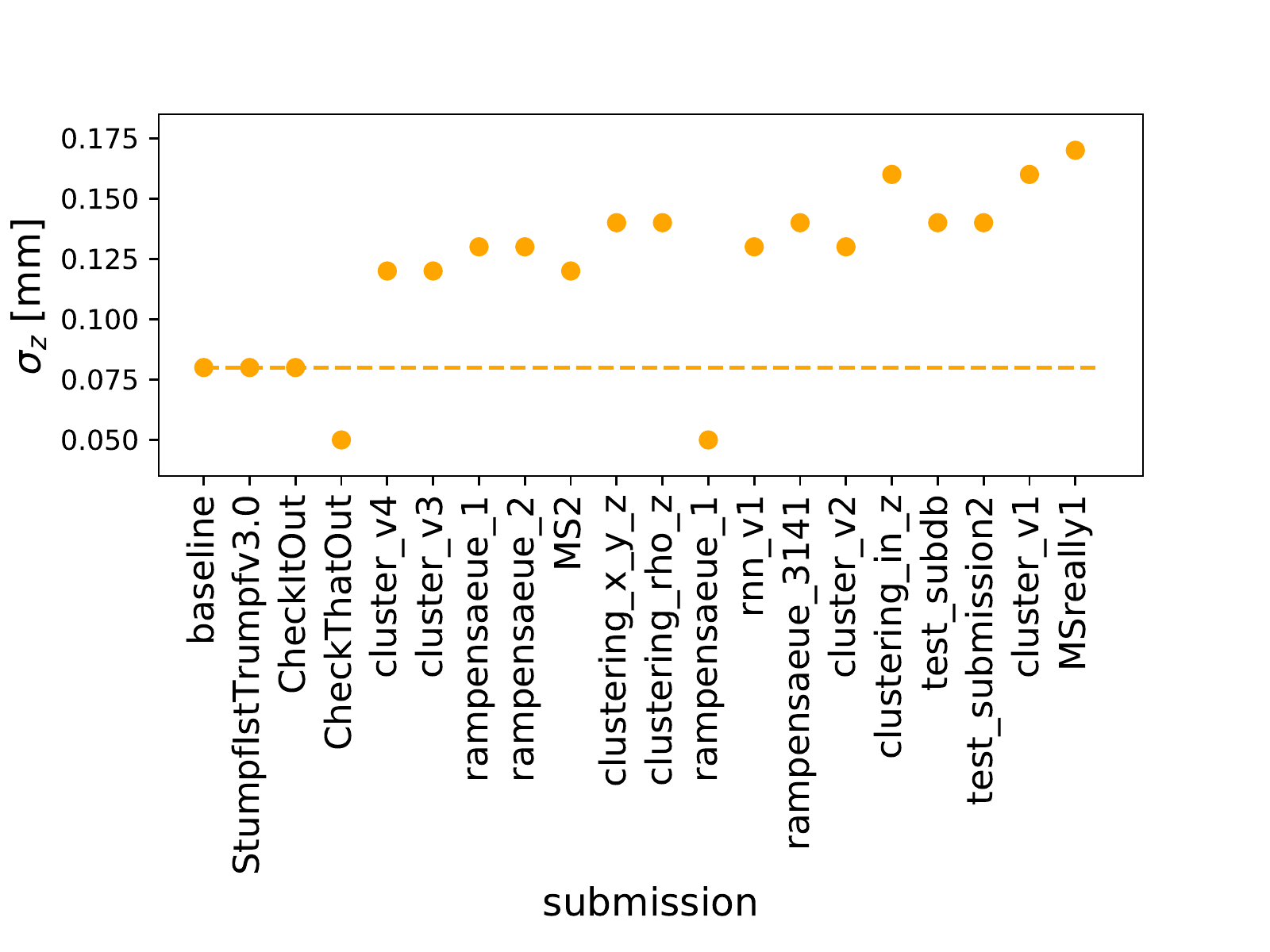}
\end{subfigure}
\caption{Comparison of the efficiency (top left), fake rate (top right), x-resolution (bottom left) and z-resolution (bottom right) of the submissions. The baseline is shown as dashed line.}
\label{fig:phys_perf}
\end{figure}

Despite the short time frame of the challenge of about three days, a number of solutions exploring various approaches were tried out.
Comparing the total scores of the submitted solutions shown in Figure~\ref{fig:tot_score} to the baseline, it can immediately seen that most of them reach a lower score, although some improve on it. This can be understood in more detail by looking at the efficiency, fake rate and x- and z-resolution of the submissions, shown in Figure~\ref{fig:phys_perf}. Most of the solutions suffer from a reduced efficiency, high fake rate and worsened z-resolution in comparison to the baseline. This indicates that the different approaches are not optimized yet due to their novelty and the relative short time available.

\begin{figure}
	\centering
	\includegraphics[height=6cm]{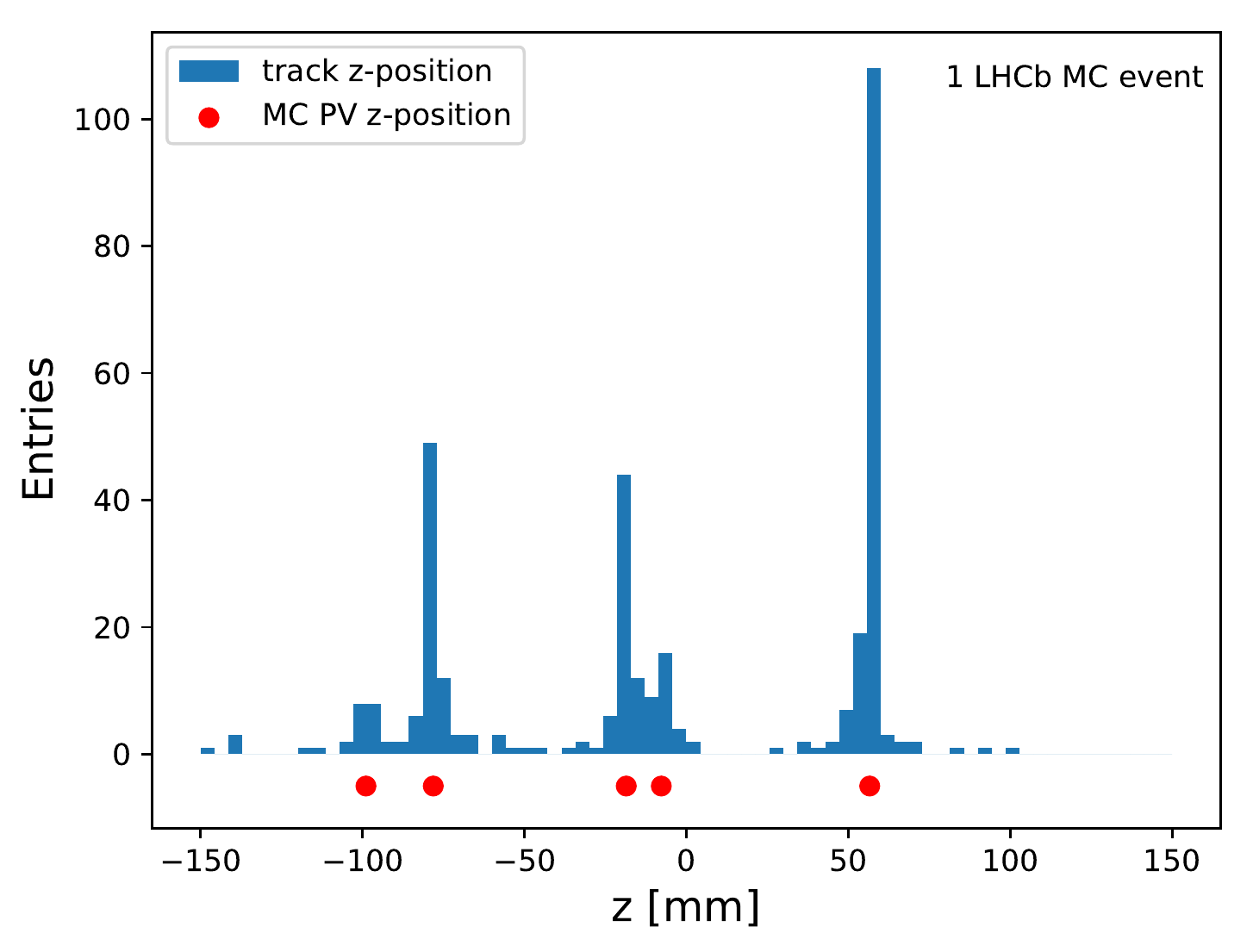}
	\caption{Histogram of z-position of tracks at point of closest approach to beamline (blue) and z-position of MC PVs in the event (red).}
	\label{fig:peaks}
\end{figure}

One striking observation is that a majority of solutions are based on a clustering algorithm. This is a well-motivated approach for this problem, as PV reconstruction can be abstracted to finding accumulations of points in a three-dimensional space, where the points are the (x,y,z) of the point of closest approach of a track to the beamline. Figure~\ref{fig:peaks} further illustrates this: It shows a histogram of the z-coordinate (at point of closest approach), where peaks, corresponding to simulated PVs shown in red, are already visible by eye.

\begin{figure}
	\centering
	\includegraphics[height=5cm]{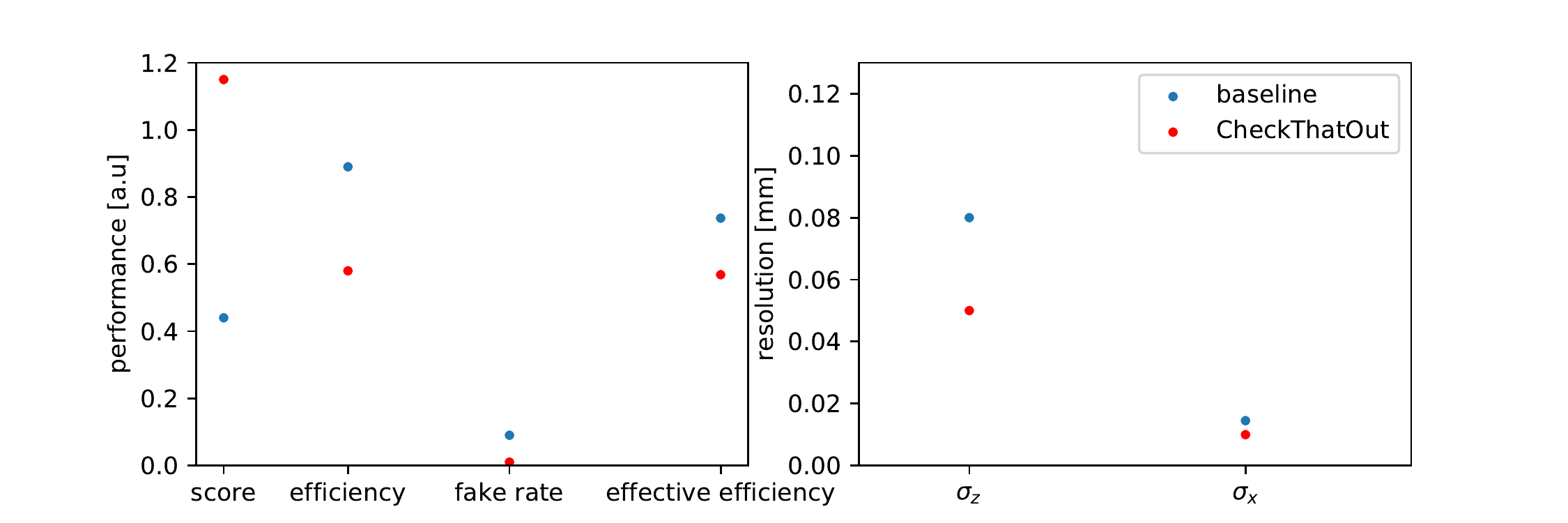}
	\caption{Comparison of a solution employing a xgboost classifier with the baseline.}
	\label{fig:xgboost}
\end{figure}

Finally, two of the highest scoring solutions are discussed in more detail. The first one, named \textit{CheckThatOut}, employs a xgboost classifier \cite{ref:xgboost} on top of the baseline solution to reduce the fake rate to almost zero, while sacrificing some efficiency. It also improves the (x,y,z)-resolution, which can be explained by the classifier mainly removing badly resolved or imprecisely reconstructed PVs. Its performance compared to the baseline is shown in Figure~\ref{fig:xgboost}.

\begin{figure}
	\centering
	\includegraphics[height=5cm]{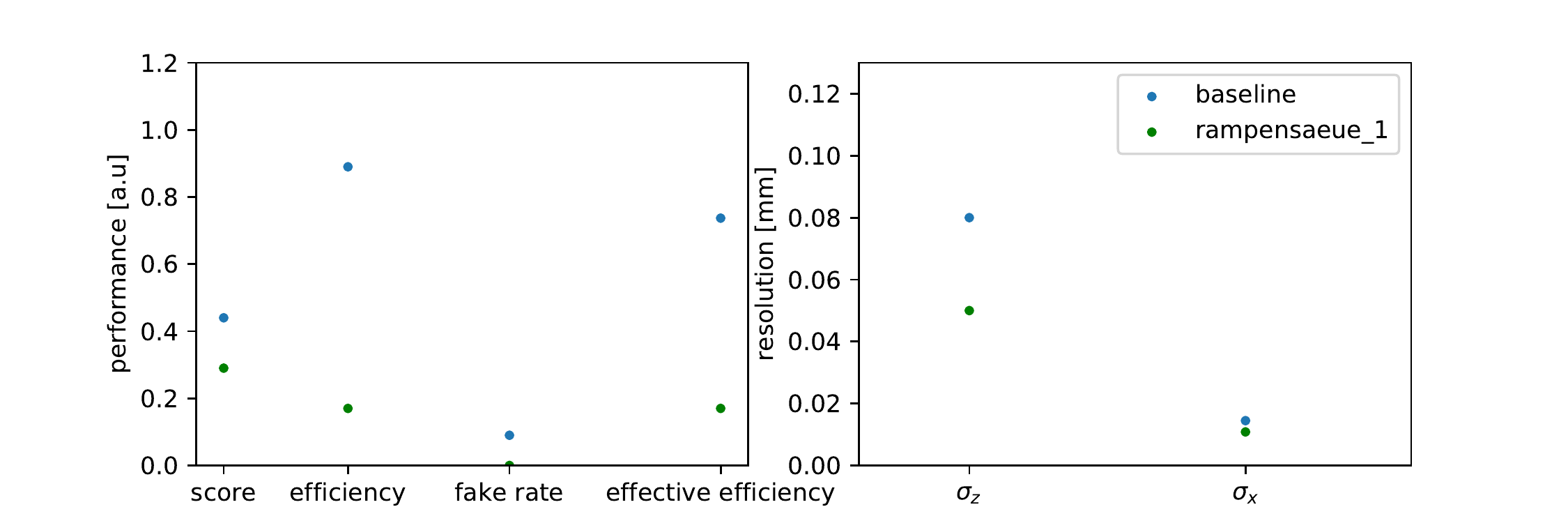}
	\caption{Comparison of a solution employing DB scan clustering with the baseline.}
	\label{fig:dbscan}
\end{figure}

The second one, called \textit{rampensaue}, uses a DB scan clustering algorithm \cite{ref:dbscan}, which yields improved position resolutions and a lower fake rate compared to the baseline, at the cost of a much lower efficiency. But in total this still results in a comparatively high score, as shown in Figure~\ref{fig:dbscan}.

\section{Conclusions}
Analyzing the solutions shows that the participants tried out different approaches, often with a similar underlying idea, with various amount of success, which should not be diminished considering the short time frame.
The challenge also highlights the importance and difficulty of choosing a scoring metric, since many parameters are of interest and have to be reduced to a single metric for the purpose of a scoreboard. This choice then affects the relative score of different solutions.
This workshops, although not without problems encountered along the way, demonstrated the utility of data challenges for the development of reconstruction algorithms for high energy physics experiments. By framing the problem in general and easy-accessible terms, participants not familiar with the LHCb experiment or particle physics at all easily could participate and try out their approaches in a short time frame and get immediate feedback on their ideas.


\Acknowledgements
I am grateful to the RAMP team for their help when preparing the challenge. I am especially indebted to Akin Kazakci and Bal\'{a}zs K\'{e}gl. Furthermore I would like to thank Vladimir Gligorov for entrusting me with this task. \\ 
I acknowledge funding from the European Research Council (ERC) under the European Union's Horizon 2020 research and innovation program under grant agreement No 724777 "RECEPT". \\
Further I acknowledge funding under the PROCOPE structural program of
the French embassies in Germany and Switzerland and of the 
European Research Council (ERC Starting Grant: PRECISION 714536).



\end{document}

%% file: econfmacros.tex
\def\beq{\begin{equation}}
\def\eeq#1{\label{#1}\end{equation}}
\def\eeqn{\end{equation}}

\def\beqa{\begin{eqnarray}}
\def\eeqa#1{\label{#1}\end{eqnarray}}
\def\eeqan{\end{eqnarray}}

\let\bar=\overbar

\def\Dslash{\not{\hbox{\kern-4pt $D$}}}
\def\dslash{\not{\hbox{\kern-2pt $\del$}}}

\def\msb{{\bar{\ssstyle M \kern -1pt S}}}

